\documentclass[twocolumn,nofootinbib,prd,aps,tightenlines]{revtex4}

\usepackage{graphicx}
\usepackage{pstricks}

{\count255=\time\divide\count255 by 60 \xdef\hourmin{\number\count255}
  \multiply\count255 by-60\advance\count255 by\time
  \xdef\hourmin{\hourmin:\ifnum\count255<10 0\fi\the\count255}}

\newcommand{\nn}{\nonumber \\ }

\providecommand{\openone}{\leavevmode\hbox{\small1\kern-3.5pt\normalsize1}}

\begin{document}

\title{
Universal Bounds for SU(3) Low Energy Constants}

\author{Vicent Mateu}
\affiliation{Departament de F\'\i sica Te\`orica, IFIC, Universitat
de Val\`encia-CSIC, Apt. Correus 22085, E-46071 Val\`encia, Spain}
  
\date{\today\quad\hourmin}

\begin{abstract}
\noindent In this paper bounds for $L_{1}$, $L_{2}$, and $L_{3}$
are obtained in chiral perturbation theory with three flavours. At
the same time we test the compatibility of this theory with axiomatic
principles. Following a recent paper we use dispersion relations to
write positivity conditions that translate into bounds for the chiral
low energy constants. As a first approach we consider the exact $SU(3)_{V}$
limit and notice that if a common mass of the order of that of the
kaon is adopted for the octet of pseudo-Goldstone bosons the bounds
have very large $\mathcal{O}(p^6)$ corrections. Once the positivity conditions are adapted
to account for different masses, we correct the previous bounds for
a physical kaon mass and find that they tighten. 
We observe an overlap between the experimentally determined region and the first principles forbidden region, in the space of parameters.
\end{abstract}

\maketitle

\section{Introduction\label{sec:Introduction}}

The pioneering idea of describing the dynamics of pions at very low
energies with an effective field theory was developed in Refs.~\cite{Wein,Gasser1}
(see also Ref.~\cite{ccwz})
and later generalized to include the $K$ and $\eta$ particles (that
is, including the $s$ quark in the light sector) in Ref.~\cite{Gasser2}.
This theory is known as chiral perturbation theory ($\chi$PT) and
its Lagrangian is organized as an infinite tower of increasing dimension
operators. Beyond the lowest order an increasing number unknown of low 
energy constants (LECs for short) must be included. The grow of LECs is even
more dramatic in the theory with three flavours [\,$SU(3)$\,] because
the Cayley-Hamilton relations are less restrictive than for the $SU(2)$
theory.

In a recent paper \cite{Manohar:2008tc} axiomatic principles such
as analyticity, unitarity, and crossing symmetry were used to
derive universal bounds for two $SU(2)$ chiral LECs. In Ref.~\cite{Ananthanarayan:1994hf}
$\chi$PT was confronted with axiomatic principles for the first time, and the method
was generalized in Ref.~\cite{Dita:1998mh}.  Some of those bounds found in Ref.~\cite{Manohar:2008tc} were already known \cite{Pennington:1994kc,Distler:2006if}, but
the most stringent conditions can only be found with the procedure
of Ref.~\cite{Manohar:2008tc}. It was also pointed out that the linear
sigma model for $m_{\sigma}\lesssim24\, m$ has a poor convergence when the $\sigma$ field is integrated out of the action, and at least corrections up to $\mathcal{O}(m_\sigma^{-6})$ must be kept to comply with the positivity bounds. 

It is the purpose of the present work
to generalize those results to the $SU(3)$ theory, and in particular
to extend the method to cover the situation of different masses [\,this
is, considering $SU(3)_{V}$ symmetry breaking\,]. In this way we will find 
out if for three flavours $\chi$PT suffers the same anomaly as the 
linear sigma model.

To our knowledge the first attempt to confront dispersion relations with three-flavour $\chi$PT to bound linear combinations of LECs was Ref.~\cite{Pham:1985}. However, in
this early work, the contribution from chiral logarithms in the $\mathcal{O}(p^4)$ amplitude
was ignored. This simplification becomes exact in the limit of an infinite number of colours, but
for a numerical analysis better results are obtained maintaining also chiral loops. In Ref.~\cite{Pham:1985} it was only possible to assert that certain linear combinations of LECs were positive and no information about the scale at which these LECs were evaluated could be
obtained.

In Ref.~\cite{Comellas:1995hq} QCD inequalities on Green functions of quark bilinear currents were used for deriving bounds on some $\chi$PT LECs. As already pointed out in Ref.~\cite{Manohar:2008tc}, we are insensitive to LECs
involving external currents, and so our results do not overlap. 

Since $\chi$PT consists of an expansion in both the external momenta
and quark masses, the coefficients of the expansion (that is, the
LECs) cannot depend on either of them. This means
that LECs do not depend on the pseudo-Goldstone bosons masses. In
other words the value of chiral LECs in our universe with $m_{s}\neq m_{u}=m_{d}$
(we will consider the isospin limit $m_{u}=m_{d}$ throughout this
paper) is the same as in ``another'' universe in which the $SU(3)_{V}$
symmetry is unbroken, $m_{s}=m_{u}=m_{d}$ . It is common
lore in the literature, for instance, to consider massless quarks for estimating
the values of some LECs, but this limit is not suitable for a dispersion relation
analysis. The most straightforward
generalization of the method used in Ref.~\cite{Manohar:2008tc} is
thus to consider the exact $SU(3)_{V}$ limit in which there are only
five independent amplitudes.

The bounds derived in this limit have two drawbacks\,: first, it
is not clear what common mass should be adopted for the degenerate octet,
what is essential to compare our bounds with the values obtained by 
fitting the experimental data (usually displayed
at the $\mu=m_{\rho}$ scale)\,; second, the results 
are not very challenging. In order to assess these two problems we will
repeat our analysis with the physical values for the $K$ and $\eta$
masses. In this case the dispersive integrals will imply positivity
conditions only under more severe conditions. Once these are addressed
the new bounds turn out to be much more restrictive, and remarkably
the central values of the fitted LEC values lie precisely on the border
dictated by axiomatic principles.

The paper is organized as follows\,: in Sec.~\ref{sec:SU(3)}
we derive the positivity conditions for the amplitudes corresponding to
the scattering of pions, kaons and etas in the $SU(3)_V$ limit
and in Sec.~\ref{sec:bounds-1}
we transform them into bounds for chiral LECs\,; in Sec.~\ref{sec:broken}
we adapt the positivity conditions to the situation of different masses
for the pseudoscalar bosons and write a new set of positivity relations\,;
in Sec.~\ref{sec:Results} we show our results\,; conclusions are
given in Sec.~\ref{sec:Conclusions}.

\section{Positivity conditions in the $\mathbf{SU(3)}$ limit\label{sec:SU(3)}}

\begin{figure}
\begin{center}
\includegraphics[width=6cm]{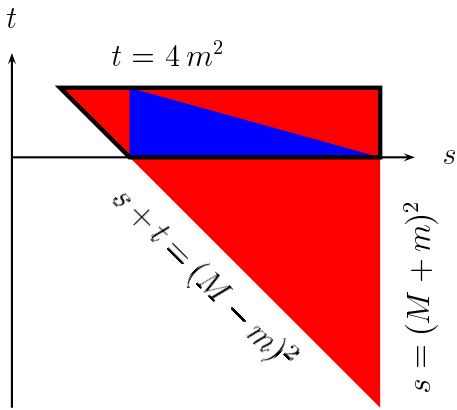}
\end{center}
\caption{Mandelstam plane for the $a+b\to a+b$ process, with $m_a=m$ and $m_b=M$ (the plot
corresponds to $m=m_\pi$ and $M=m_K$). The small (blue) triangle in the center is the Mandelstam triangle. The big triangle (red and blue area) is the region free from singularities. The region bounded by the thick black line corresponds to the area $\mathcal{A}$ in which the positivity conditions are satisfied.\label{fig:pipi-diagram} }
\end{figure}

In this section we straightforwardly apply the methods of Ref.~\cite{Manohar:2008tc}
to the pseudoscalar-pseudoscalar scattering processes. As a first approach we consider
the $m_{u}=m_{d}=m_{s}$ limit, and so the pseudoscalar octet has
a common mass which we denote by $m$. The detailed derivation of
the positivity conditions can be found in Ref~\cite{Manohar:2008tc}
and will be only sketched here. Further details will be given in Sec.~\ref{sec:broken}
when we consider flavour symmetry breaking. Since there is no
lighter particle in the QCD spectrum than the pseudo-Goldstone bosons
(pGs for short) the analytic structure is fully dictated by two-pGs
intermediate states. Much as happened in $\pi\,\pi$ scattering, the branch
cuts emerge for $s,\, t,\, u>4\, m^{2}$ what is equivalent to say
that the amplitude is analytic in the Mandelstam plane for $s,\, t\le4\, m^{2}$
and $s+t\ge0$. This result relies on perturbation theory to all orders \cite{Eden-book}, but using solely axiomatic principles it can be shown (Ref.~\cite{Martin:1965jj}) that they are at least valid in the interval $-\,14\,m^2\ge t \ge 4\,m^2$, which is enough for our purposes.

In the limit we are considering the QCD Lagrangian exhibits an exact
$SU(3)_{V}$ symmetry. Then particles are classified according to
the different irreducible representations of this group (\emph{e.g.} pGs
belong to the real octet representation) and
the Wigner-Eckart theorem drastically reduces the number of independent amplitudes to
six. To see this we simply need to look at the Clebsch-Gordan decomposition
of the direct product of two octets\,:\begin{equation}
8\otimes8\,=\,27\oplus10\oplus10^{*}\oplus8_{1}\oplus8_{2}\oplus1\,.\label{eq:C-G}\end{equation}
On the other hand one can find a representation analogous to the Chew-Mandelstam
in $SU(3)$~%
\footnote{One must remember the $SU(3)$ identity $3\left(d^{abe}d^{cde}+d^{ace}d^{bde}+d^{ade}d^{bce}\right)=\delta^{ab}\delta^{cd}+\delta^{ac}\delta^{bd}+\delta^{ad}\delta^{bc}$
to make sure that the basis of tensors is minimal. One can also add
four more structures of the type $f^{abe}d^{cde}$, but they clash
after imposing crossing symmetry.%
} \begin{eqnarray}
T(ab\to cd) & = & A_{1}(s,t,u)\,\delta^{ab}\delta^{cd}+A_{2}(s,t,u)\,\delta^{ac}\delta^{bd}\nn
&&\,+\,A_{3}(s,t,u)\,\delta^{ad}\delta^{bc}\,+\, B_{1}(s,t,u)\, d^{abe}d^{cde}\nn
&&\,+\,B_{2}(s,t,u)\, d^{ace}d^{bde}\,.\label{eq:cartan-SU(3)}\end{eqnarray}
Since Eq.~(\ref{eq:cartan-SU(3)}) has only five independent amplitudes
there must be one identity relating the amplitudes of Eq.~(\ref{eq:C-G}).
In fact crossing symmetry forces $T_{10}(s,t)=T_{10^{*}}(s,t)$, making
Eqs.~(\ref{eq:C-G}) and (\ref{eq:cartan-SU(3)}) compatible. We
also expect crossing symmetry to further reduce the number of independent
functions. In case of having $r$ irreducible representation amplitudes
[\,$r=3$ for $SU(2)$ and $r=6$ for $SU(3)$\,] crossing symmetry
implies that there are only $\frac{r}{3}$ independent functions. This is easy to understand\,:
the $r$ irreducible functions $T^I$ $I=1,\ldots r$ translate into $3\,r$ degrees of freedom
$T^I(s,t)$, $T^I(t,s)$, and $T^I(4\,m^2-s-t,t)$ corresponding to the $s$-, $t$-, and $u$- crossed
channels, respectively. Crossing symmetry implies $2\,r$ restrictions, since it relates the 
$s$-channel amplitudes with the $t$- and $u$-channel ones (\,$2\,r$ relations). 
So we end up with $r$
independent degrees of freedom, which is equivalent to $\frac{r}{3}$ independent functions. 
So in $\pi\,\pi$ scattering there is only one independent function (\emph{e.g.}
the Chew-Mandelstam coordinate $A$) while 
we are left
with two independent functions. All in all for $SU(3)$ we can write the following
crossing relation
\begin{eqnarray}
T^{\, I}(s,t)&=& C_{u}^{II^{\prime}}T^{\, I^{\prime}}(u,t)\,,\quad C_{u}^{II^{\prime}}C_{u}^{I^{\prime}J}\,=\,\delta_{IJ}\,, \nonumber\\[3pt]
C_{u}&=&\left(\begin{array}{ccccc}
\frac{7}{40} & \phantom{-\,}\frac{1}{6} & \phantom{-\,}\frac{1}{5} & \phantom{-\,}\frac{1}{3} & \phantom{-\,}\frac{1}{8}\\[3pt]
\frac{9}{40} & \phantom{-\,}\frac{1}{2} & \phantom{-\,}\frac{2}{5} & \phantom{-\,}0 & -\,\frac{1}{8}\\[3pt]
\frac{27}{40} & \phantom{-\,}1 & -\,\frac{3}{10} & -\,\frac{1}{2} & \phantom{-\,}\frac{1}{8}\\[3pt]
\frac{9}{8} & \phantom{-\,}0 & -\,\frac{1}{2} & \phantom{-\,}\frac{1}{2} & -\,\frac{1}{8}\\[3pt]
\frac{27}{8} & -\,\frac{5}{2} & \phantom{-\,}1 & -\,1 & \phantom{-\,}\frac{1}{8}\end{array}\right)\,,
\end{eqnarray}
and analogously for $T^{\, I}(t,s)$. We use $I,\, J=27$, $10$, $8_{1}$, $8_{2}$, $1$
to denote the irreducible amplitudes of Eq.~(\ref{eq:C-G}), not
to be confused with isospin. Exchanging the order of the initial or
final particle amounts to change $t$ by $4\, m^{2}-s-t$. Under this
operation the amplitudes $I=1,\,8_{1}\,,27$ remain invariant and
the rest change sign. 

Following Ref.~\cite{Manohar:2008tc} we can write the following twice-subtracted
dispersion relation
\begin{eqnarray}
\frac{\mathrm{d}^{2}}{\mathrm{d}s^{2}}T^{I}(s,t)&=&\frac{2}{\pi}\int_{4m^{2}}^{\infty}\mathrm{d}x\left[\,\frac{\delta^{II'}}{(x-s)^{3}}\right. \nn&&\,+\left. \frac{C_{u}^{II'}}{(x-u)^{3}}\right]\mathrm{Im}\, T^{I'}(x+i\epsilon,t)\,,\label{eq:disp_final}
\end{eqnarray}
wherever $(s,t)$ makes the amplitude analytic, that is $t\le4\, m^{2}$,
$s+t\ge0$ and if $s>4\, m^{2}$ considering $s\to s+i\,\epsilon$, corresponding to the Feynman
prescription for propagators. Clearly, if we restrict ourselves to $s<4\, m^{2}$ and $s+t>0$, both denominators in Eq.~(\ref{eq:disp_final}) are positive.
As shown in Ref.~\cite{Manohar:2008tc}, for several linear combinations $\sum a_{I}\,T^{I}$ with $a_{I}\ge0$, $\sum a_{I}\,C_u^{IJ}\,T_J\equiv\sum_J b_J \,T_J$ with $b_J=\sum_I  a_{I}\, C_{u}^{IJ} \ge0$.
These have a positive imaginary part along the integral for $t>0$, corresponding to physical
processes with equal initial and final states. Of course, many different
processes are related by $SU(3)$ symmetry and need to be considered
only once. If a process can be expressed as a linear combination of
other processes with positive coefficients it cannot be more restrictive
than the processes separately, so it will be discarded. With all that
we obtain the following set of positivity conditions\,:
\begin{widetext}
\begin{eqnarray}
\frac{\mathrm{d}^{2}}{\mathrm{d}s^{2}}\,T(\pi^{+}\pi^{+}\to\pi^{+}\pi^{+})[\,(s,t)\in\mathcal{A}\,]\ge0\,,\quad &  &\quad \frac{\mathrm{d}^{2}}{\mathrm{d}s^{2}}\,T(\pi^{0}\pi^{0}\to\pi^{0}\pi^{0})[\,(s,t)\in\mathcal{A}\,]\ge0\,,\nn
\frac{\mathrm{d}^{2}}{\mathrm{d}s^{2}}\,T(\pi^{+}\pi^{0}\to\pi^{+}\pi^{0})[\,(s,t)\in\mathcal{A}\,]\ge0\,,\quad&&\quad \frac{\mathrm{d}^{2}}{\mathrm{d}s^{2}}\,T(\eta\,\pi\to\eta\,\pi)[\,(s,t)\in\mathcal{A}\,]\ge0\,,\nn
\frac{\mathrm{d}^{2}}{\mathrm{d}s^{2}}\,T(K\,\eta\to K\,\eta)[\,(s,t)\in\mathcal{A}\,]\ge0\,,\quad&& \quad  \frac{\mathrm{d}^{2}}{\mathrm{d}s^{2}}\,T(K\pi^{+}\to K\pi^{+})[\,(s,t)\in\mathcal{A}\,]\ge0\,,\label{eq:processes}
\end{eqnarray}
\end{widetext}
where $\mathcal{A}$ is the closed region of the Mandelstam plane
defined by $0\le t\le4\, m^{2}$, $s\le4\, m^{2}$, $s+t\ge0$ 
(see Fig.~\ref{fig:pipi-diagram}). Equation~(\ref{eq:processes})
corresponds to the following linear combinations of irreducible amplitudes
\begin{eqnarray}
&&\frac{27}{20}\,T_{27}+\frac{1}{5}\,T_{8_{1}}+\frac{1}{8}\,T_{1}\,,\quad  \frac{1}{2}\,T_{27}+\frac{1}{3}\,T_{8_{2}}+\frac{1}{5}\,T_{10}\,,\nn &&
\frac{3}{10}\,T_{27}+\frac{1}{5}\,T_{8_{1}}+\frac{1}{2}\,T_{10}\,, \qquad \frac{1}{2}\,T_{27}+\frac{1}{2}\,T_{10}\,, 
\nn &&  T_{27}\,,\qquad  \frac{9}{20}\,T_{27}+\frac{1}{20}\,T_{8_{1}}+\frac{1}{4}\,T_{8_{2}}+\frac{1}{4}\,T_{10}\,,\end{eqnarray}
respectively.

\section{Bounds on $\mathbf{L_{1},\; L_{2}}$ and $\mathbf{L_{3}}$.\label{sec:bounds-1}}

It is straightforward now to convert the positivity conditions in
Eq.~(\ref{eq:processes}) into bounds for chiral LECs, since the
energy domain $\mathcal{A}$ is well inside the convergence radius
of $\chi$PT. We simply plug into Eq.~(\ref{eq:processes}) the $\mathcal{O}(p^{4})$
$\chi$PT prediction [\,the $\mathcal{O}(p^{2})$ prediction vanishes
when acting with two derivatives\,] for the different amplitudes and
seek the most stringent point in $\mathcal{A}$. These amplitudes
can be found in the literature but are collected and very nicely displayed
in Ref.~\cite{GomezNicola:2001as}, which we follow. Upon the second
derivative they only depend on three LECs\,: $L_{1},\; L_{2}$, and
$L_{3}$. At one loop the amplitudes explicitly depend on the chiral
renormalization scale $\mu$, but it is in fact canceled by the implicit
$\mu$ dependence of the chiral LECs. We will adopt the value $\mu=m$
that greatly simplifies the expressions (as it is the only energy
scale in the process). So we will get our bounds for $L_{1}$ and
$L_{2}$ evaluated at that energy scale ($L_{3}$ does not get renormalized
and thus it is $\mu$ independent). Our bounds have the following
general expression\begin{equation}
\alpha_{1i}\, L_{1}^{r}(m)+\alpha_{2i}\, L_{2}^{r}(m)+\alpha_{3i}\, L_{3}^{r}\ge f_{i}[\,(s,t)\in\mathcal{A}\,]\Bigr|_{\mathrm{max}}\,,\end{equation}
where $f_{i}$ are functions obtained by isolating the LECs of the
second derivative of the amplitude\,: it contains chiral logarithms
and constant LEC-independent terms. For all processes the maximum is achieved for
$t=4m^2$. For the processes $\pi^{+}\pi^{+}\to\pi^{+}\pi^{+}$ and $K\,\pi^{+}\to K\,\pi^{+}$
the minima are found for $s=1.3684\, m^{2}$ and $s=1.2593\, m^{2}$, respectively. For the rest of the processes it is found for $s=0$.

If we are to compare our theoretical
bounds with the fitted values we need to fix the common mass $m$
to a physical value. The most conservative value is of course the
pion mass $m_{\pi}$, since it is the lightest particle in the octet,
but in principle any value low enough not to compromise the chiral
expansion is equally good. We will adopt the two extreme values $m_{\pi}$
and $m_{K}$ for our analysis. The results are shown in Table~\ref{tab:results}.

If we consider the more realistic case of $m_{s}\neq m_{u}=m_{d}$
and use the physical value for the $\pi$ and $K$ states~%
\footnote{In our analysis we will assume the Gell-Mann-Okubo formula for the
masses\,: $m_{\eta}^{2}=\frac{4}{3}m_{K}^{2}-\frac{1}{3}m_{\pi}^{2}.$%
} the choice of $m$ is absolutely transparent. This is discussed in
the next section.

\section{Symmetry corrections to the bounds\label{sec:broken}}

The first effect showing up when considering $m_{\pi}<m_{K}$ is that
for several processes the unitarity branch cut might occur before
reaching the physical threshold. This, as we discuss next, spoils
the positivity condition. 

Let us first obtain the analytic triangle for the present situation.
We will consider only processes with equal initial and final states
$a+b\to a+b$, of masses $m_{a}=M$ and $m_{b}=m$ ($M\ge m$), since
this ensures that the imaginary part of the partial wave amplitudes
is positive. If the lowest mass intermediate state in that process
is $c+d$, the amplitude is analytic for $s\le(m_{c}+m_{d})^{2}$.
Analogously from the crossed channels we will obtain $t\le(m_{e}+m_{f})^{2}$
and $s+t\ge2\,(m^{2}+M^{2})-(m_{g}+m_{h})^{2}$. Of course the maximum
[\,minimum\,] value for these three thresholds are $(m+M)^{2}$, $4\, m^{2}$,
and $(M-m)^{2}$, respectively. Then the dispersion relation reads
(now we directly consider physical processes)
\begin{eqnarray}
\frac{\mathrm{d}^{2}}{\mathrm{d}s^{2}}\,T(s,t)&=&\frac{2}{\pi}\int_{(m_{c}+m_{d})^{2}}^{\infty}\mathrm{d}x\,\frac{\mathrm{Im}\, T(x+i\epsilon,t)}{(x-s)^{3}}\nn&+&\frac{2}{\pi}\int_{(m_{g}+m_{h})^{2}}^{\infty}\mathrm{d}x\,\frac{\mathrm{Im}\, T_{u}(x+i\epsilon,t)}{(x-u)^{3}}\,,
\end{eqnarray}
wherever the amplitude is analytic. Using only axiomatic principles \cite{Martin:1965jj}
it can be shown that for $K\,\pi$ and $\eta\,\pi$ scattering, dispersion relations are
valid at least for $-\,32.76\,m^2\ge t\ge 4\,m^2$ and $-\,37.85\,m^2\ge t\ge 4\,m^2$,
respectively. Here $T_{u}$ is the amplitude
corresponding to the $u$-channel $a+\bar{b}\to a+\bar{b}$, which
has, of course, equal initial and final states, too. Both denominators
are positive as far as $s\le(m_{c}+m_{d})^{2}$ and $s+t\ge2\,(m^{2}+M^{2})-(m_{g}+m_{h})^{2}$,
and so up to this point there is nothing compromising the positivity
condition. But still we have to make sure that the imaginary part
remains positive along the two cuts. Expanding the amplitude $T$ (and
also $T_u$) in partial waves we get
\begin{equation}
T(s,t)=\!\sum_{\ell=0}^{\infty}(2\,\ell+1)f_{\ell}(s)P_{\ell}\!\!\left[1+\frac{s\, t}{(s+m^{2}-M^{2})^{2}-4m^{2}s}\right]\!,
\end{equation}
with $\mathrm{Im}\, f_{\ell}(s)\,=\, s\,\beta(s)\,\sigma_{\ell}(s)\,\theta\,[\,s-(m_{c}+m_{d})^{2}\,]\,\ge\,0$
and with $\theta\,[\,s-(m_{g}+m_{h})^{2}\,]$ for the $u$-channel. So
for getting a positive imaginary part each $P_{\ell}$ must be positive
along the corresponding cuts. Since $P_{\ell}(z)>1$ for $z>1$ for
all $\ell$ it is enough to require
\begin{equation}
\frac{s\, t}{(s+m^{2}-M^{2})^{2}-4 m^{2}s}\ge0\quad\mathrm{for}\; s\ge\left\lbrace \begin{array}{c}(m_{c}+m_{d})^{2}\\[3pt](m_{g}+m_{h})^{2}\end{array}\right. .\label{eq:spoiler}
\end{equation}
Since for $s\to\infty$ Eq.~(\ref{eq:spoiler}) tends to $t/s$ then
we must require $t>0$. Then for positive $t$ Eq.~(\ref{eq:spoiler})
is only satisfied if $(M-m)^{2}\ge s\ge(M+m)^{2}$. Thus if either
$(m_{c}+m_{d})$ [\,or $(m_{g}+m_{h})$\,] is less than $(M+m)$ the
imaginary part between $(m_{c}+m_{d})$ [\,or $(m_{g}+m_{h})$\,] and
the physical threshold could turn negative, making the positivity
condition invalid.

Summarising, \emph{the positivity conditions hold for processes of
the type $a+b\to a+b$ such that the lightest pair of particles that
can arise off the scattering $a+b$ is precisely $a+b$, and analogously
for $a+\bar{b}$}. Or in other words, for processes with equal initial and final states
such that the imaginary part of the $s$- and $u$-channels starts at their physical production  threshold. 
Moreover, the positivity condition is satisfied
in the closed area of the Mandelstam plane $\mathcal{A}$ defined
by $0\geq t \geq 4\,m^2$, $s\leq (M+m)^2$ and $s+t\geq(M-m)^2$ (see Fig.~\ref{fig:pipi-diagram}).
As an additional bonus for breaking $SU(3)_{V}$ we have many independent
amplitudes that are no longer related by symmetry. The final set of positivity
conditions reads\,:
\begin{widetext}
\begin{eqnarray}
\frac{\mathrm{d}^{2}}{\mathrm{d}s^{2}}\,T(\pi^{+}\pi^{+}\to\pi^{+}\pi^{+})[\,(s,t)\in\mathcal{A}_{\,\pi}\,]\ge0\,,\quad && \quad \frac{\mathrm{d}^{2}}{\mathrm{d}s^{2}}\,T(\pi^{0}\pi^{0}\to\pi^{0}\pi^{0})[\,(s,t)\in\mathcal{A}_{\,\pi}\,]\ge0\,,\nn
\frac{\mathrm{d}^{2}}{\mathrm{d}s^{2}}\,T(\pi^{+}\pi^{0}\to\pi^{+}\pi^{0})[\,(s,t)\in\mathcal{A}_{\,\pi}\,]\ge0\,,\quad
&& \quad 
\frac{\mathrm{d}^{2}}{\mathrm{d}s^{2}}\,T(\eta\,\pi\to\eta\,\pi)[\,(s,t)\in\mathcal{A}_{\,\eta}\,]\ge0\,,\; \nn
\frac{\mathrm{d}^{2}}{\mathrm{d}s^{2}}\,T(K\,\pi^{+}\to K\,\pi^{+})[\,(s,t)\in\mathcal{A}_{\,K}\,]\ge0\,,\quad&&
\end{eqnarray}
\end{widetext}
where of course, the area $\mathcal{A}$ depends on each specific process. There are more processes satisfying the conditions stated
above, but they give a less stringent bound for the same linear combination
of LECs and so we will not show them. Again all minima are found at $t=4\,m^2$. For the $\pi^{+}\pi^{0}$, $\eta\,\pi$, and $K\,\pi^{+}$ processes the minima are achieved for
$s=1.14384\, m^{2}$, $s=16.0027\, m^{2}$, and $s=4.78\, m^{2}$, respectively. For the remaining
two processes, it is found at $s=0$.

\section{Results\label{sec:Results}}

\renewcommand{\arraystretch}{1.5}

\begin{table*}
\begin{center}\begin{tabular}{|c|c|c|c|c|c|}
\hline
Process&
$10^{3}\,\alpha_{i}\,L^{i}(m_{\rho})$&
Fit to exp.&
Bound $m=m_{\pi}$&
Bound $m=m_{K}$ &
Bound $m_\pi\neq m_{\pi}$ \tabularnewline
\hline
$\pi^{0}\pi^{0}$&
$2\, L_{1}^{r}+2\, L_{2}^{r}+L_{3}$&
$\phantom{}-\,0.03\,(1.03)\pm0.5\phantom{5}$&
$\ge\,\,-\,3.88\pm0.20$&
$\ge0.68\pm2.50$&
$\ge\,\,-\,3.87\pm3.00$\tabularnewline
$\pi^{+}\pi^{0}$&
$L_{2}^{r}$&
$\phantom{0-}\,0.73\,(1.59)\pm0.12$&
$\ge\,\,-\,1.30\pm0.20$&
$\ge0.22\pm2.50$&
$\ge\,\,-\,1.10\pm3.00$\tabularnewline
$\pi^{+}\pi^{+}$&
$2\, L_{1}^{r}+3\, L_{2}^{r}+L_{3}$&
$\phantom{0-}\,0.70\,(2.62)\pm0.6\phantom{5}$&
$\ge\,\,-\,4.88\pm0.20$&
$\ge1.20\pm2.50$&
$\ge\,\,-\,4.29\pm3.00$\tabularnewline
$K\,\eta$&
$12\, L_{2}^{r}+L_{3}$&
$\phantom{-}\,6.41\,(16.17)\pm1.5\phantom{5}$&
$\ge-\,15.99\pm0.20$&
$\ge2.24\pm2.50$&
-\tabularnewline
$\eta\,\pi$&
$3\, L_{2}^{r}+L_{3}$&
$\;\,-\,0.16\,(1.86)\pm0.5\phantom{5}$&
$\ge\,\,-\,3.64\pm0.20$&
$\ge0.92\pm2.50$&
$\ge\,\,-\,0.15\pm3.00$\tabularnewline
$K^{+}\pi^{+}$&
$4\, L_{2}^{r}+L_{3}$&
$\phantom{-0}\,0.57\,(3.45)\pm0.6\phantom{5}$&
$\ge\,\,-\,4.70\pm0.20$&
$\ge1.38\pm2.50$&
$\ge-\,14.75\pm3.00$\tabularnewline
\hline
\end{tabular}\end{center}

\caption{Experimental values for linear combinations of the LECs and their
bounds. In the third column, for the values displayed in brackets we use the fitted values of the LECs when using an $\mathcal{O}(p^4)$ $\chi$PT theoretical prediction.\label{tab:results}}
\end{table*}

In this section we discuss the bounds obtained for the different linear
combinations of chiral LECs, and compare them with the values obtained
by fitting observables to the experimental data. In Ref.~\cite{Amoros:2001cp}
those values are given at the $\mu=m_{\rho}$ scale, so we will run
our bounds to this scale to compare. The running equation for these LECs reads
\begin{eqnarray}
&&L_{i}(\mu_{1})\,-\,L_{i}(\mu_{2})\,=\,-\,\frac{\Gamma_{i}}{16\pi^{2}}\log\left(\frac{\mu_{1}}{\mu_{2}}\right)\,,\nn
&&\Gamma_{1}\,=\,\frac{3}{32}\,,\quad\Gamma_{2}\,=\,\frac{3}{16}\,,
\end{eqnarray}
and the values at the different scales are \cite{Amoros:2001cp}
\begin{eqnarray}
L_{1}^{r}(m_{\rho})&=&(0.43\,[\,0.38\,]\,\pm\,0.12)\times10^{-3}\,,\nn
L_{2}^{r}(m_{\rho})&=&(0.73\,[\,1.59\,]\,\pm\,0.12)\times10^{-3}\,,\nonumber\\
L_{3}&=&(-\,2.35\,[\,2.91]\,\pm\,0.37)\times10^{-3}\,.\label{eq:L-values}
\end{eqnarray}
Those values were obtained from a fit to the available experimental data taking as theoretical input the $\mathcal{O}(p^6)$ $\chi$PT prediction. Since in our analysis we are using the $\mathcal{O}(p^4)$ amplitude it is instructive to compare our bounds with the values of the LECs obtained by fitting the $\mathcal{O}(p^4)$ $\chi$PT amplitude to the same data. Those can be found in Ref.~\cite{Amoros:2001cp} as well, and are displayed in Eq.~(\ref{eq:L-values}) in brackets. 

A very important issue is to estimate the error committed by truncating the amplitude at
$\mathcal{O}(p^4)$. The $\mathcal{O}(p^6)$ amplitude is divided into three pieces\,: two-loop terms, that only depend on masses\,; one-loop terms, that depend on several $\mathcal{O}(p^4)$ LECs\,; and tree-level terms, that depend on $\mathcal{O}(p^6)$ LECs. For the symmetric
analysis the error can be estimated as in Ref.~\cite{Manohar:2008tc}, that is, adopting as an educated guess 3 times the corrections due to double chiral logarithms. When assuming
$m=m_\pi$ the bounds are not very stringent and the errors are rather small\,; experimental values are well within the bounds. However, for $m=m_K$ the central values of the bounds greatly increase (that is, bounds tighten) and some experimental values apparently violate the bounds. But at the same time errors get multiplied by a factor of 12. 
Thence the validiy of the chiral expansion is not compromised.

For the symmetry breaking analysis the error cannot be estimated so
straightforwardly. It is expected that the main corrections come from chiral LECs multiplied
by the kaon mass. The $\mathcal{O}(p^6)$ computation of the $\pi\,\pi$ scattering amplitude in three-flavour $\chi$PT was performed in Ref.~\cite{Bijnens:2004eu}, and the $K\,\pi$ scattering at the same order can be found in Ref.~\cite{Kpi-two-loop}. We will adopt as an educated guess the correction due to the $\mathcal{O}(p^6)$ LECs, that is the $\mathcal{O}(p^6)$ tree-level piece. Unfortunately the  $\mathcal{O}(p^6)$ LECs are unknown, so we will use the estimate given in Refs.~\cite{Kpi-two-loop,Bijnens:2004eu}, obtained by resonance saturation. In addition, to be more conservative, we will assume a common error for all the channels, the biggest of these, which is $3.0$. This error is very large, of the same order as that of the symmetric analysis with $m=m_K$.

For the three $\pi\,\pi$ scattering processes we do not see large deviations of the corrected bounds (they increase around  20\%). However the estimated error due to higher order corrections greatly enhances due to terms proportional to the kaon mass. So we can conclude that the symmetric analysis is most convenient for these relations. Incidentally experimental values satisfy these three bounds. For $K\,\pi$ scattering the corrected bound is much worse. However for $\eta\,\pi$ scattering the increase of the corrected bound is great\,: 139\%. In fact the experimentally fitted value is partially in conflict with the bound, but since the error of the bound is quite large, the validity of $\chi$PT is not compromised. 

The bounds compare better to the values of the LECs obtained from an $\mathcal{O}(p^4)$ fit. It is quite easy to understand this. The bounds are to a large extent dominated by the value of $L_2$, since in the corresponding linear combinations it always appears multiplied by large coefficients (see second column of Table~\ref{tab:results}). In Eq.~(\ref{eq:L-values}) we see that the value of $L_2$ in the $\mathcal{O}(p^4)$ fit is twice as big as in the  $\mathcal{O}(p^6)$.

Results are displayed in Table~\ref{tab:results}. In the first column
we show which process is rendering each bound and in the second the
corresponding linear combination of LECs, in the third column we display
the corresponding linear combinations of the experimentally fitted values from the $\mathcal{O}(p^6)$ fit, and in brackets when using the values from the $\mathcal{O}(p^4)$ fit\,;
in the fourth and fifth columns we display the bounds for the symmetric analysis
assuming $m=m_\pi$ and $m=m_K$, respectively\,; in the last column we give 
the bounds obtained for broken $SU(3)_{V}$ symmetry.

\section{Conclusions\label{sec:Conclusions}}

As demonstrated in Ref.~\cite{Manohar:2008tc} the combination of
effective field theories and axiomatic principles turns out to be
a powerful tool for disentangling some properties of nonperturbative
phenomena. The latter render model independent positivity conditions
that yield bounds on the LECs of the former.

We apply this program to $\chi$PT with three flavours and find bounds
for $L_{1}$, $L_{2}$, and $L_{3}$. When the exact $SU(3)_{V}$ limit
is considered the bounds become badly convergent if the common mass $m$
for the multiplet of pGs is of the order of $m_{K}$ (albeit they converge well
for $m=m_{\pi}$). When the actual values for the pion and kaon
masses are employed the bounds become more stringent and, in fact, in one case the experimentally fitted values are partially in contradiction with the central value of
the bound. However, for this process the $\mathcal{O}(p^6)$ corrections are very large and so there is no contradiction.

\section*{Acknowledgments}
V.~M. is grateful to professors A.~Pich and A.~V.~Manohar for careful reading of the
manuscript, and to professor J.~R.~Pelaez for helpful correspondence.
The work of V.~Mateu is supported by a FPU contract (MEC). This work
has been supported in part by the EU MRTN-CT-2006-035482 (FLAVIAnet),
by MEC (Spain) under Grant No. FPA2004-00996 and by Generalitat Valenciana
under Grant No. GVACOMP2007-156.

\end{document}